\documentclass[twocolumn,pre,aps,showpacs,amsmath,amssymb]{revtex4-1}
\usepackage{graphicx}
\usepackage{bm}
\usepackage{epstopdf}
\begin{document}

\title{Nonequilibrium Thermodynamics of Restricted Boltzmann Machines}
\author{Domingos S. P. Salazar}
\email[]{salazar.domingos@gmail.com}
\affiliation{Unidade de Educa\c{c}\~{a}o a Dist\^{a}ncia e Tecnologia, Universidade Federal Rural de Pernambuco, Recife, Pernambuco 52171-900 Brazil}

%\date{\today}

\begin{abstract}
In this work, we analyze the nonequilibrium thermodynamics of a class of neural networks known as Restricted Boltzmann Machines (RBMs) in the context of unsupervised learning. We show how the network is described as a discrete Markov process and how the detailed balance condition and the Maxwell-Boltzmann equilibrium distribution are sufficient conditions for a complete thermodynamics description, including nonequilibrium fluctuation theorems. Numerical simulations in a fully trained RBM are performed and the heat exchange fluctuation theorem is verified with excellent agreement to the theory. We observe how the contrastive divergence functional, mostly used in unsupervised learning of RBMs, is closely related to nonequilibrium thermodynamic quantities. We also use the framework to interpret the estimation of the partition function of RBMs with the Annealed Importance Sampling method from a thermodynamics standpoint. Finally, we argue that unsupervised learning of RBMs is equivalent to a work protocol in a system driven by the laws of thermodynamics in the absence of labeled data.
\end{abstract}
\maketitle

\section{Introduction}
\label{intro}

%General Intro of NNs and learning.
Neural networks learn from noisy environments by adjusting its internal configuration (or weights) in order to map input variables into known outputs (or labels). This type of learning is known as supervised and it requires a reasonable volume of labeled data. When this condition is met, supervised learning becomes extremely effective in a variety of applications, specially with deep architectures consisting on multiple layers of neurons \cite{DeepReview,HintonScience2006}. More recently, a stochastic thermodynamic analysis of a supervised learning rule was successfully developed \cite{SeifertPRL2017}, enhancing the understanding of supervised learning efficiency. Despite of the importance of supervised learning, most of the biologic systems learning tasks are likely to happen unsupervised, taking place in the absence of labeled data \cite{Barlow1989}. The general process of synaptic plasticity resulting in learning representations of the world from unlabeled sensory data seems essential in the quest for understanding intelligence \cite{Dayan2005,RevModPhys2006}. Such important role in biologic systems poses a question whether unsupervised learning is also ruled by some fundamental laws such as thermodynamics.

% Deep Learning (mostly supervised)
In addition to the desirable ability of learning representations, unsupervised learning also had major importance in the origins of deep learning \cite{Hinton2006}. Although shallow artificial neural networks (ANNs) have been used for a long time, the use of multiple layers of neurons in ANNs, the so called deep learning, had practical applications just recently \cite{DeepReview}. Deep learning was put forward with the introduction of the contrastive divergence (CD) learning algorithm \cite{Hinton2002} to pre train a specific type of network, the Restricted Boltzmann Machine (RBM), as building blocks of deep architectures \cite{HintonDBN}. Those new ideas influenced older supervised successful algorithms \cite{Lecun1990} to be recast into speech \cite{Hintonspeech} and image recognition \cite{AlexNet} problems with new available data and computational power resulting in extraordinary performance. Deep learning applications are now used from high energy physics \cite{Atlas1,Atlas2} and phase transitions \cite{PhaseTrans2016} to genomics \cite{Genomics} and gaming \cite{Alphago}. 
%Some explanation to the outstanding results of deep learning have been proposed. In image recognition, the local structure of the data and simple physical constraints (such as translation invariance) favor architectures based on local transformations \cite{DLworks1}. 
%Another explanation refers to the flattening of manifold-shaped data in deep architectures \cite{DLworks2}.
% Unsupervised needs Physical meaning.
%The success of very deep and data intensive supervised algorithms questioned the real necessity of the original unsupervised approach of pre training the models. A recent review \cite{DeepReview} shows concern about this issue and puts unsupervised learning as a potential area of research that may unlock breakthroughs in machine learning due to the vast amount of unlabeled data available. The resemblance to the learning process of living organisms, such as vision [Ref] and speech [Ref] also corroborates with the importance of this type of learning.

%We start by RBMs because its physically motivated.
In order to understand some fundamental laws of unsupervised learning, we study a type of stochastic neural network known as Restricted Boltzmann Machine (RBM) \cite{Hinton2002,RBM1}. Beyond their important role in the development of deep learning, RBMs are closely related to physical systems \cite{RBMPRB2016}. The RBM has two layers (called visible and hidden layers) of binary units. There are connections between any neurons from different layers, but neurons from the same layer are not connected, therefore the network forms a bipartite graph. The RBM has a scalar energy function associated to each state of the network and the probability of finding a state is given by the Maxwell-Boltzmann (MB) distribution. The network is generative in the sense that it can be used to randomly create data (visible layer) from a given configuration of the hidden layer. The unsupervised training of the RBM requires a systematic adjustment of its weights until it is able to generate data from the training distribution with some accuracy. In this case, we say the network has learned the data distribution in an unsupervised way.

%Training RBMs
The contrastive divergence (CD) algorithm \cite{Hinton2002} allows unsupervised learning of RBMs with a very simple learning rule for updating its weights. During the training process, the RBM learns internal (hidden) representations of the input (visible) data until it is able to generate random outputs that resembles the original data. The CD learning algorithm in RBMs has been used in image recognition \cite{HintonDBN} but also applied to learn other general data distributions, such as the Ising model \cite{RBMPRB2016}. Some adaptations to CD have been proposed \cite{PCD}, but they remain based on the Gibbs sampling procedure, which is the rule used to randomly generate one layer of the network as a stochastic function of the other layer.
%Ref Therm learning thermodynamics observables.

%In this paper
In this paper, we show that unsupervised learning of RBMs is somehow driven by thermodynamics. We start by introducing a physical motivation for RBMs by considering the Gibbs sampling procedure as the systems discrete time dynamics. In this case, physical observables, such as the energy, fluctuate randomly in discrete time, akin to other systems from continuous stochastic thermodynamics~\cite{Seifert2008,Sekimoto2010, Seifert2012}. As a consequence, the system behaves as a Markov chain and satisfies the detailed balance condition. Suitable definitions of thermodynamic work and heat are adapted from a framework \cite{Crooks1998} for discrete Markov chains. As the Gibbs sampling dynamics allows the RBM to exchange heat and perform work, the system behaves in accordance to the first law of thermodynamics. As expected from the framework used, we obtain the Crooks Fluctuation Theorem (CFT)~\cite{Crooks1998} and the underlying second law of thermodynamics. We also show how the RBM, initially prepared in equilibrium at temperature $T_1$, obeys the heat exchange fluctuation theorem (XFT)~\cite{Jar2004} when placed in contact a reservoir of different temperature $T_2$, with excellent agreement to numerical simulations. Then, we analyze the contrastive divergence (CD) learning functional within the nonequilibrium thermodynamics framework and rewrite it in terms of physical quantities. Finally, we use the concepts presented in the paper to interpret Annealed Importance Sampling (AIS), a known method for estimating the partition function of RBMs, in the light of stochastic thermodynamics concepts.

The paper is organized as follows. Section II introduces mathematical properties of RBMs and defines the thermodynamic observables resulting in the first law. Section III treats the derivation of nonequilibrium fluctuation theorems for the system compared to numerical simulations and obtains the second law. Section IV describes the unsupervised CD learning written in terms of nonequilibrium thermodynamic quantities. Section V uses the framework to interpret AIS method for estimating the partition function. Section VI contains conclusion and perspectives on unsupervised learning understood as a thermodynamic process.

\section{First law of thermodynamics in RBMs}
\label{maths}
In this section, we review the formalism of Restricted Boltzmann Machines (RBMs) in the context of a discrete stochastic process. Then, we allow the weights (parameters of the RBM) to change in time, which in turn leads to a natural definition of thermodynamic observables (heat and work) and the underlying first law.

%\subsection{Restricted Boltzmann Machines}
The structure of RBMs \cite{RBM1,Hinton2002} is composed by two layers of neurons (or units) with binary states. The visible layer ($m$ units) is fully connected to the hidden layer ($n$ units), however there are not connections between neurons in the same layer. The state $s$ of the network is determined by a pair of vectors formed by the states of the visible ($v$) and hidden ($h$) neurons, $s=(v,h)$, where $v=\{v_i\}$, $i=1,...,m$, and $h=\{h_j\}$, $j=1,...,n$. The neurons $v_i$ and $h_j$ assume values $0$ or $1$. In a given configuration $\lambda$, the energy of a given state $s$ is defined:
\begin{equation}\label{Energy}
    E(s,\lambda)=-\sum_{i=1}^{m}a_{i}v_{i}-\sum_{j=1}^{n}b_{j}h_{j}-\sum_{i,j=1}^{m,n}v_{i}w_{ij}h_{j},
\end{equation}
where $s=(v,h)$ is a state and $\lambda=\{a_i,b_j,w_{ij}\}$ represents a configuration of weights (or parameters). The model also assigns a probability for each state of the network that depends only on its energy:
\begin{equation}\label{probability}
   p_{\lambda}(s)=\frac{1}{Z(\beta,\lambda)}e^{-\beta E(s,\lambda)},
\end{equation}
where the partition function, $Z(\beta,\lambda)=\sum_{s} e^{-\beta E(s,\lambda)}$, is the sum of the Boltzmann factor over all possible states of the network, assuring the probability adds up to 1. Notice that (\ref{probability}) is the Maxwell-Boltzmann (MB) probability distribution from statistical mechanics. We have deliberately included the parameter $\beta$ representing the inverse temperature ($\beta=1/T$, for $k_B=1$). Most original RBM formulations set $\beta=1$, because $\beta$ is usually kept constant during simulations. Recently, temperature has been introduced as a parameter in the temperature based RBM for a variety of purposes \cite{Bengio2010,TRBM}. The properties of the temperature based RBM remain unchanged with the introduction of a constant $\beta$, since the parameter could be rescaled in the original weights $\lambda$ by a simple transformation, $\beta\lambda\rightarrow\lambda$. However, the introduction of temperature allows the notion of different thermal reservoirs, which is a central motivation for the heat exchange fluctuation theorem (XFT) discussed in the next section.

A very useful property of RBMs is the independence of the neurons from the same layer. Using (\ref{Energy}) and (\ref{probability}) it can be deduced \cite{HintonTrends2007} the conditional probability of finding a hidden (visible) unit given a visible (hidden) vector:
\begin{equation}\label{conditional1}
   p_{\lambda}(h_j=1|v)=\sigma(\beta b_j+\beta \sum_{i}v_iw_{ij}),
\end{equation}
\begin{equation}\label{conditional2}
   p_{\lambda}(v_i=1|h)=\sigma(\beta a_i+\beta \sum_{j}h_jw_{ij}),
\end{equation}
where $\sigma(x)=1/(1+exp(-x))$ is the sigmoid function. The subscript $\lambda$ is explicitly written for clarity, since they will be adjusted during the learning process. This property above makes it possible to numerically estimate sample averages easily, which are used during training the parameters $\lambda$.

Simulations on RMBs use (\ref{conditional1}) and (\ref{conditional2}) to generate a layer based on the opposite layer as a Markov chain. This is called the Gibbs sampling \cite{Hinton2002} and it works as if the dynamics of a RBM understood as a discrete time Markov chain. 
%In this case, using Bayes theorem, the conditional probability of finding a state $s'=(v',h')$ given a known state $s=(v,h)$ is given by:
%\begin{equation}\label{transition}
 %  p_{\lambda}(s'|s)=p_{\lambda}(v',h'|v,h)=p_{\lambda}(v'|h)p_{\lambda}(h'|v).
%\end{equation}
The conditional probabilities in the identity above can be written in terms of (\ref{conditional1}) and (\ref{conditional2}) as
\begin{eqnarray}
\label{transitionvh}
p_{\lambda}(v|h)=\prod_i^m p_{\lambda}(v_i|h), \\
p_{\lambda}(h|v)=\prod_i^n p_{\lambda}(h_j|v).
\end{eqnarray}
Although the probability (\ref{transitionvh}) is a conditional probability defined from the MB distribution (\ref{probability}), the Gibbs sampling dynamics assigns it to the transition probability of a single step, $p^{(1)}_{\lambda}(s\rightarrow s')$, from a state $s=(v,h)$ to a final state $s=(v',h')$ in the discrete stochastic process: 
\begin{equation}\label{dynamics}
p^{(1)}_{\lambda}(s\rightarrow s')\equiv p_{\lambda}(v'|h)p_{\lambda}(h'|v'),
\end{equation}
with $p_{\lambda}(v'|h)$ and $p_{\lambda}(h'|v')$ defined in (\ref{transitionvh}). The statistical dependence of the layers are depicted in Fig.1. This equivalence is the starting point of the thermodynamic analysis, since it defines a stochastic dynamics that encodes an arrow of time. In this case, it is clear that the dynamic process (\ref{dynamics}) is a Markov chain, since by definition the probability of finding state $s_K$ at time step $K$ depends only on the previous state.
%\begin{equation}\label{markov}
%   p_{\lambda}(s_K|s_{K-1},...,s_0)=p_{\lambda}(s_{K}|s_{K-1}).
%\end{equation}
In other words, the dynamics does not have a memory of previous states of the chain. It is also essential to notice that the Gibbs sampling dynamical process defined in (\ref{dynamics}) satisfies detailed balance condition. For $K=1$ step, detailed balance reads for a constant $\lambda$:
\begin{equation}\label{detailedbalance1}
   \frac{p_{\lambda}^{(1)}(s\rightarrow s')}{p_{\lambda}^{(1)}(s'\rightarrow s)}
   =\frac{p_{\lambda}(v'|h)p_{\lambda}(h'|v')}{p_{\lambda}(v|h)p_{\lambda}(h|v')}=\frac{p_{\lambda}(s')}{p_{\lambda}(s)},
\end{equation}
where the last identity was obtained using Bayes theorem. For multiple steps, $K>1$, notice that the transition probability may be written in terms of the one step transitions. For simplicity, we consider a constant $\lambda$, which is the relevant case for the heat exchange fluctuation theorem:
\begin{equation}\label{GibbsK}
   p_{\lambda}^{(K)}(s\rightarrow s')
   =\sum_{s_1,...,s_{K-2}}\prod_{i=0}^{K-1}p^{(1)}_{\lambda}(s_i\rightarrow s_{i+1}),
\end{equation}
where $p_{\lambda}^{(K)}(s\rightarrow s')$ is the transition probability of state $s$ to state $s'$ after $K$ steps in the dynamics, $s_0=s$ and $s'=s_K$. Upon using (\ref{detailedbalance1}) in (\ref{GibbsK}), one obtains the detailed balance condition also for $K$ steps
\begin{equation}\label{detailedbalance}
   \frac{p_{\lambda}^{(K)}(s\rightarrow s')}{p_{\lambda}^{(K)}(s'\rightarrow s)}
   =\frac{p_{\lambda}(s')}{p_{\lambda}(s)},
\end{equation}

\begin{figure}[ht]
\includegraphics[width=3.0 in]{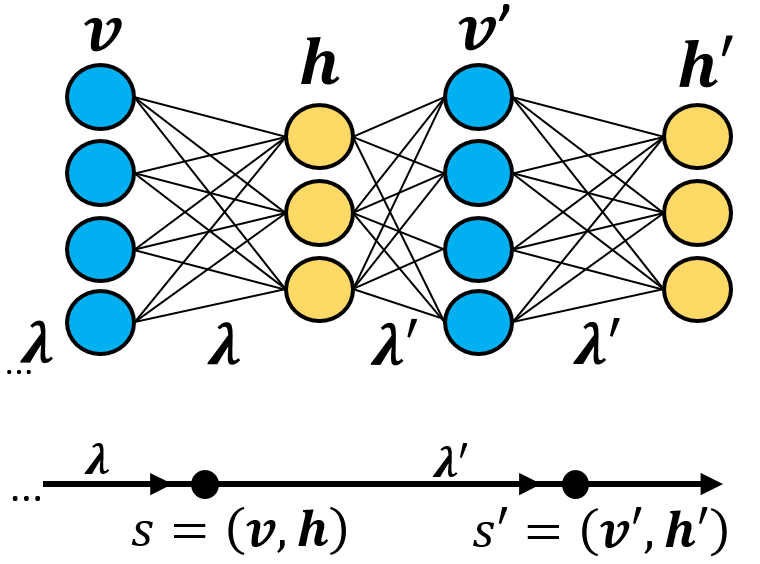}
\caption{(Color online) The figure shows a RBM with $m=4$ units in the visible layer (blue), $v$, and $n=3$ units in the hidden layer (yellow), $h$. Notice that neurons from the same layer (same color) are not connected. The example shows the RBM performing a single step of the Markov chain, initially at state, $s=(v,h)$, generated with weights $\lambda$, transitioning into another state, $s'=(v',h')$, generated with weights $\lambda'$.}
\label{fig1}
\end{figure}
where $p_{\lambda}(s)$ and $p_{\lambda}(s')$ are MB distributions (\ref{probability}). As discussed in the next section, detailed balance plays a major role in the derivation of nonequilibrium fluctuation theorems (FTs).

In general, during the discrete steps of the dynamics depicted in Fig.1, one could allow the weights $\lambda$ to be adjusted as a function of time. Actually, the process of learning in RBMs (and other networks) is a type of weight adjustment and it can be done in many ways. In all type of learning rules, there will be an iterative change of parameters configuration, from $\lambda=\{a_i,b_j,w_{ij}\}$ to $\lambda'=\{a_i',b_j',w_{ij}'\}$, where $\lambda$ may be understood as an external set of controlled parameters. There is also noise from the stochastic (Gibbs sampling) dynamics itself. More precisely, defining the energy $E(s_k,\lambda_k)$, of the configurations $s_k$ generated with $\lambda_k$, given by (\ref{Energy}), the variation $\Delta E$ during a sequence of $K$ steps, $\Sigma=(s_0,...,s_K)$, is given by $\Delta E = E(s_{K},\lambda_K)-E(s_0,\lambda_0)$. This variation can be conveniently written as a contribution of two factors as pointed in \cite{Crooks1998} for discrete Markov chains:
\begin{equation}\label{Heat}
Q=\sum_{k=0}^{K-1} E(s_{k+1},\lambda_{k+1})-E(s_k,\lambda_{k+1}),
\end{equation}
\begin{equation}\label{Work}
W=\sum_{k=0}^{K-1} E(s_k,\lambda_{k+1})-E(s_k,\lambda_k),
\end{equation}
which can be understood as the heat and work for the trajectory $\Sigma=(s_0,...,s_K)$. Definitions above result in the first law of thermodynamics for RBMs, since $\Delta E = W + Q$. From the specific form the energy (\ref{Energy}) in RBMs one gets complete expressions for work and heat in terms of the neurons values ($v,h$) and the network configuration $\lambda$. Notice that during learning processes, there are changes in the weights from $\lambda_k$ to $\lambda_{k+1}$, which allows the work to be different from zero in (\ref{Work}). Alternatively, heat accounts for the energy variation due to stochastic change of states $s_k\rightarrow s_{k+1}$, even with the same configuration $\lambda$, in analogy with the thermodynamics observable also found in other stochastic systems \cite{Sekimoto2010}.

\section{Fluctuation Theorems and the Second Law}
\label{FTs}
In this section, we use general properties of the dynamics of the RBMs as Markov chains to derive known fluctuation theorems and the second law of thermodynamics. There are complete reviews of fluctuation theorems (FTs) in Markov systems with continuous time dynamics \cite{Seifert2012,Harris2007}. Here we explore FTs in the discrete time dynamics observed in RBMs.

\subsection{Crooks Fluctuation Theorem and the Second Law}

In a discrete Markov chain, the system undergoes a given trajectory, $\Sigma=(s_0,...,s_K)$, with the configuration being adjusted in a controllable protocol, $\Lambda=(\lambda_0,...,\lambda_K)$. The work, $W$, defined in (\ref{Work}) is a random variable that depends on the trajectory. In this case, the Crooks Fluctuation Theorem (CFT) \cite{Crooks1999} states a property for the probability density function of the random variable $W$ as the identity: 
\begin{equation}\label{CFT}
\frac{P_{s_0\rightarrow s_K}(W)}{P_{s_0\rightarrow s_0}(-W)}=e^{\beta(W-\Delta F)},
\end{equation}
where $P_{s_0\rightarrow s_K}(W)$ is the probability of finding the thermodynamic work over all trajectories going from state $s_0$ to state $s_K$. The variation of the free energy is defined in terms of the partition function, $\Delta F=-T \log (Z(\beta,\lambda_K)/Z(\beta,\lambda_0))$, for $k_B=1$. The theorem is valid when both the initial and final distributions are the equilibrium distribution. For RBMs, the derivation of CFT is easily adapted from the original formulation \cite{Crooks1999}, since the result was originally introduced for discrete Markov chains satisfying detailed balance, which is the case of this paper. The slight difference comes from the multidimensional control parameter $\lambda$. Therefore, we will keep the calculations brief and refer to the original when needed.

Defining the backwards trajectory as $\gamma'=(s_K,...,s_0)$, we can write from (\ref{dynamics}) and the Markov property:
\begin{equation}\label{CFTdetailed}
\frac{P(\gamma)}{P(\gamma')}=\frac{p_{eq}(s_0)}{p_{eq}(s_K)}\prod_{k=0}^{K-1}\frac{p_{\lambda_{k+1}}(s_{k+1}|s_k)}{p_{\lambda_{k+1}}(s_k|s_{k+1})},
\end{equation}
where $P(\gamma)$ is the probability of the trajectory $\gamma$. After rearranging the factors above and using detailed balance (\ref{detailedbalance}), one gets:
\begin{equation}\label{CFTdetailed2}
\frac{P(\gamma)}{P(\gamma')}=
\frac{p_{eq}(s_0)}{p_{\lambda_0}(s_0)}
\frac{p_{\lambda_K}(s_K)}{p_{eq}(s_K)}
\prod_{k=0}^{K-1}\frac{p_{\lambda_k}(s_k)}{p_{\lambda_{k+1}}(s_k)}.
\end{equation}
Finally, using the explicit form of the equilibrium distributions in (\ref{detailedbalance}) leads to
\begin{equation}\label{CFTdetailed2}
\frac{P(\gamma)}{P(\gamma')}=
\frac{p_{eq}(s_0)}{p_{\lambda_0}(s_0)}
\frac{p_{\lambda_K}(s_K)}{p_{eq}(s_K)}
e^{\beta(W-\Delta F)},
\end{equation}
with $W$ defined in (\ref{Work}) as the work of the forward trajectory $\gamma$. Considering the initial and final distributions to be equilibrium distributions (\ref{probability}) and summing over all possible trajectories with the same work leads to the identity (\ref{CFT}).

A consequence of CFT (\ref{CFT}) is the Jarzynski equality (JE) \cite{Jar1997}:
\begin{equation}\label{JE}
\langle e^{-\beta W} \rangle = e^{-\beta \Delta F}.
\end{equation}
where the ensemble average above is taken over all possible trajectories also starting from configurations $\lambda_0$ to $\lambda_K$. Jensen's inequality, $\langle exp(x) \rangle \geq exp\langle x \rangle$, applied in (\ref{JE}) results in $\langle W \rangle \geq \Delta F$, which is the the second law of thermodynamics. Actually, by defining the Shannon entropy as
\begin{equation}\label{Entropy}
S(\beta,\lambda)=-\sum_{s}p_{\lambda}(s)\log{p_{\lambda}(s)},
\end{equation}
and using (\ref{probability}), one gets the expression for the entropy variation from configurations $\lambda_0$ to $\lambda_K$ (with constant temperature):
\begin{equation}\label{EntropyT}
\Delta S=\beta\langle \Delta E \rangle +\log(Z(\beta,\lambda_K)/Z(\beta,\lambda_0)).
\end{equation}
Now using (\ref{Heat}) and (\ref{Work}), the expression for the entropy finally gets the form
\begin{equation}\label{EntropyVar}
\Delta S=\beta\langle Q \rangle + \beta\langle W \rangle - \beta \Delta F \geq \beta\langle Q \rangle,
\end{equation}
where the inequality follows from $\langle W \rangle - \Delta F \geq 0$, which represents the irreversible work in fine time processes \cite{Sekimoto2010}. Expression above is a common statement of the second law of thermodynamics.

\subsection{Heat Exchange Fluctuation Theorem}
\label{XFTsec}
In the absence of work, the energy variation in the RBM is totally due to heat exchange. When approaching equilibrium, the system's energy variation after a single discrete time step is expected to approach zero on average. However, when the RBM is prepared with a given temperature $T_1$ and then placed in thermal contact with a reservoir with a different temperature $T_2$, there will be a nonequilibrium fluctuation for the heat $Q$ random variable (\ref{Heat}). The heat exchange fluctuation theorem (XFT) \cite{Jar2004} states a identity for the nonequilibrium heat probability:
\begin{equation}\label{XFT}
\frac{P(Q)}{P(-Q)}=e^{Q(\beta_1-\beta_2)},
\end{equation}
for $Q$ the heat transferred to the RBM, where $\beta_1$ and $\beta_2$ are the inverse temperatures of reservoirs 1 and 2, respectively. The identity holds for any number of steps $K$ in (\ref{Heat}), where we have adapted the identity to our sign notation for the heat. The original derivation uses a small coupling between the two systems in consideration \cite{Jar2004}, as well as continuous time dynamics. In the case of RBMs, as expected in other stochastic systems \cite{Seifert2012}, we show that XFT also follows from the discrete Markov dynamics with detailed balance as an exact result, without further assumptions on the magnitude of the coupling. We start by noticing that $\Delta E = Q$, in the absence of work. The probability, $P(\Delta E)$, of finding the energy variation, $\Delta E$, after any number $K$ of steps is given in terms of the joint probability of states:
\begin{equation}\label{XFT1}
P^{(K)}(\Delta E)=\sum_{s,s'} p_2^{(K)}(s\rightarrow s')p_1(s)\delta(E'-E-\Delta E),
\end{equation}
where $s$ and $s'$ are the initial and final states with energies $E=E(s)$ and $E'=E(s')$. For clarity. the probabilities $p_2$ and $p_1$ are generated from (\ref{GibbsK}) and (\ref{probability}), using $T_2$ and $T_1$, respectively, with the same constant $\lambda$ in both cases (the $\lambda$ subscript was omitted for simplicity). The function $\delta$ is defined as $\delta(x)=1$, if $x=0$, and $\delta(x)=0$ otherwise. Replacing the MB distribution $p_1(s)$ using (\ref{probability}) leads to
%Now we use Bayes theorem to write $p(s',s)=p(s'|s)p(s)$, where $p(s)$ is the MB probability distribution (\ref{probability}) with temperature $T_1$, leading to

\begin{equation}
P^{(K)}(\Delta E)=\sum_{s,s'} p_2^{(K)}(s\rightarrow s')\frac{e^{-\beta_1 E}}{Z(\beta_1,\lambda)}\delta(E'-E-\Delta E).
\end{equation}
The expression above can be rearranged easily after the introduction of the equilibrium distribution $p_1(s')$
\begin{equation}
P^{(K)}=e^{\beta_1 \Delta E}\sum_{s,s'} p_2^{(K)}(s\rightarrow s')p_1(s')\delta(E'-E-\Delta E),
\end{equation}
and after exchanging the summation variables ($s,s'$), one gets
\begin{equation}
P^{(K)}(\Delta E)=e^{\beta_1 \Delta E}\sum_{s',s} p_2^{(K)}(s'\rightarrow s)p_1(s)\delta(E'-E+\Delta E).
\end{equation}
Applying the detailed balance (\ref{detailedbalance}) in the transition probability $p_2^{(K)}(s'\rightarrow s)$ above leads to
\begin{eqnarray}
\label{XFTlast}
P^{(K)}(\Delta E)=e^{(\beta_1-\beta_2)\Delta E}\times
\\
\times\sum_{s',s} p_2^{(K)}(s\rightarrow s')p_1(s)\delta(E'-E+\Delta E),
\end{eqnarray}
where last double sum can be identified as $P^{K}(-\Delta E)$ from definition (\ref{XFT1}). Finally, equation (\ref{XFTlast}) results in the identity 
\begin{equation}
\label{XFTlast2}
P^{(K)}(\Delta E)=e^{(\beta_1-\beta_2)\Delta E}P^{(K)}(-\Delta E),
\end{equation}
for any $K$, which is the original XFT identity (\ref{XFT}).
\begin{figure}[ht]
\includegraphics[width=3.0 in]{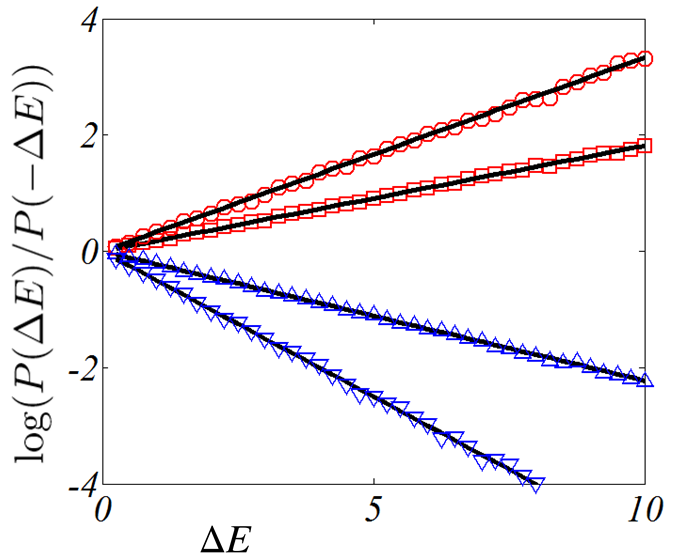}
\caption{(Color online) The figure shows the heat exchange fluctuation theorem (XFT) for a Restricted Boltzmann Machine (RBM) trained over a set of handwritten digits (MNIST), with $m=784$ units in the visible layer ($v$) and $n=500$ units in the hidden layer ($h$). Theoretical predictions are the solid lines. The initial temperature is $T_1=1$ and the final temperatures are $T_2=1.2$ ($\circ$), $T_2=1.1$ ($\square$) for heating (in red) and $T_2=0.9$ ($\triangle$), $T_2=0.8$ ($\triangledown$) for cooling examples (in blue). The energy difference pdf, $P(\Delta E)$, is obtained in the nonequilibrium situation of a single update ($K=1$) of the discrete Markov chain with $T_2$, from an ensemble of $2\cdot10^7$ RBMs initially prepared at thermal equilibrium with $T_1$.}
\label{fig2}
\end{figure}

In order to verity the XFT numerically, a simulation was implemented using a known RBM architecture ($m=784,n=500$) \cite{AISforRBM} for the task of image recognition of hand written digits (MNITS) \cite{MNIST}. Details on the unsupervised training of the RBM are given in Appendix (\ref{Appendix1}). After training the configuration $\lambda$, an ensemble of $N=2\cdot10^7$ RBMs, with the same $\lambda$, was put in thermal equilibrium with the temperature $T_1=1$. The equilibrium was prepared after a Markov chain Monte Carlo simulation for several steps ($K>100$) from an initial random state and $T_1=1$. Finally, we take the final state $s$ of each RBMs, supposedly in equilibrium, with energy $E=E(s)$, and perform a single step in the dynamics ($K=1$) with a different temperature $T_2$ and same configuration $\lambda$. This procedure results in the state $s'$ and energy $E'=E(s')$. The energy variation, $\Delta E = E'-E$, is used to compute the numerical pdf from $N=2\cdot10^7$ RMBs and the ratio $P(\Delta E)/P(-\Delta E)$ is evaluated and displayed in FIG. \ref{fig2} for different final temperatures $T_2=\{0.8,0.9,1.1,1.2\}$. In all cases, the XFT  predictions (\ref{XFT}) for the nonequilibrium case of $K=1$ step are remarkably consistent with the simulations. Notice that the ascending lines, $T_2=\{1.2, 1.1\}$, suggest that the heating process ($T_2>T_1$) favors a positive variation of the energy ($P(\Delta E)/P(-\Delta E)>1$), as expected. Alternatively, the descending lines, $T_2=\{0.8,0.9\}$, represent cooling processes ($T_2<T_1$) for which the energy of the RBM is expected to decrease ($P(\Delta E)/P(-\Delta E)<1$). It is important to notice that the results derived above (\ref{XFTlast2}) are true for any parameter configuration $\lambda$, including a biased model fully trained over a data set as presented. 

The property of heat exchange has been observed experimentally for different physical systems \cite{XFTexp01}. The derivation presented above for a discrete Markov chain relying on detailed balance is general and it suits well the formalism of RBMs presented in this paper. The same approach and numerical simulation setup could possibly be applied in other learning systems with deep architectures \cite{HintonDBN}, where the RBMs are used as building blocks.

\section{Unsupervised learning as thermodynamic process}
In this section, we explore the unsupervised learning process of contrastive divergence (CD) in the context of nonequilibrium thermodynamics. First, we review the necessary notation of the CD. Then, we analyze the relation between the algorithm and thermodynamics.

\subsection{Contrastive Divergence (CD)}\label{ssCD1}

The contrastive divergence (CD) algorithm \cite{Hinton2002} is one of the most successful unsupervised learning rules for RBMs. It works by updating the weights of a RBM iteratively so it better generates a given data distribution. Due to its simplicity and speed, several applications of RBM as generative models became possible \cite{HintonScience2006}. In this section, we analyze CD in the framework of stochastic thermodynamics introduced in this paper.
The algorithm is motivated by the optimization of the log likelihood function, $L(\lambda,D)$, over a training data set $D=\{v_i\}^{N}_{i=1}$ of $m$ dimensional vectors, $v_i$, defined as
\begin{equation}\label{loglike}
    L(\lambda,D)=\sum_{i=1}^{N}\log p_{\lambda}(v_i),
\end{equation}
where $p_{\lambda}(v_i)$ the is the observed probability (\ref{probability}) of $v_i$ given by the model with configuration $\lambda$. A perfect model would reproduce training data exactly, thus $p_{\lambda}(v_i)=1$ for all $i$, resulting in $L(\lambda,D)=0$. But this ideal situation is not reachable in real data sets. Typically, one would adjust the weights, $\lambda=\{a_i,b_i,h_{ij}\}=\{\theta\}$, of the generative model iteratively to maximize the log likelihood (\ref{loglike}), using a stochastic gradient ascent (SGA) approach \cite{Hinton2002}. In every iteration $\tau$, the Maximum Likelihood (ML) learning increments each parameter, $\theta_\tau$, from the set $\lambda_\tau$, as
\begin{equation}\label{SGA}
    \theta_{\tau+1} = \theta_{\tau} + \eta\cdot\partial_{\theta} L(\lambda,D)\rvert_{\lambda_\tau},
\end{equation}
where $\partial_{\theta}=\partial/\partial \theta$ is a short notation for the partial derivative, the constant $\eta$ is a positive learning rate and $L(\lambda,D)$ is taken from (\ref{loglike}). Upon replacing (\ref{probability}) in (\ref{loglike}), the RBM gets a simple form for the expression (\ref{SGA}), where the increments of the parameters may be easily represented as averages of the energy function (\ref{Energy}):
\begin{equation}\label{derivative}
    \partial_{\theta} L(\lambda,D)=
    -\beta(\langle \partial_{\theta} E\rangle_{D}-\langle \partial_{\theta}E\rangle_{\lambda}),
\end{equation}
where the partial derivatives are immediate due to the linear dependence of $E(s,\lambda)$ taken from (\ref{Energy}) for each parameter $\{\theta\}=\{a_i,b_i,w_{ij}\}$. The expression  $\langle f(v,h) \rangle_{D}$ represents the average of a function of the state $s=(v,h)$ over the training data $D$:
\begin{equation}\label{averagedata}
    \langle f(v,h) \rangle_{D} = \sum_{v\in D,h}p_{D}(v)p_{\lambda}(h|v)f(v,h),
\end{equation}
with $p_{D}(v)$ representing the relative frequency of $v\in D$ and $p_{\lambda}(h|v)$ given in (\ref{transitionvh}). Similarly, the value $\langle f(v,h) \rangle_{\lambda}$ represents the average of the function $f(v,h)$ as evaluated by the RBM with parameters $\lambda$,
\begin{equation}\label{averagemodel}
    \langle f(v,h) \rangle_{\lambda}=\sum_{v,h}p_{\lambda}(v,h)f(v,h),
\end{equation}
where $p_{\lambda}(v,h)=p_{\lambda}(s=(v,h))$, given by (\ref{probability}). Inserting (\ref{derivative}) in (\ref{SGA}) leads to the increments for each parameter $\theta$ of the RBM:
\begin{align}
\label{CDapp}
\nonumber
\Delta a_i&= -\eta \beta (\langle v_i \rangle_D - \langle v_i \rangle_\lambda),\\
\Delta b_j&= -\eta \beta (\langle h_j \rangle_D - \langle h_j \rangle_\lambda),\\ \nonumber
\Delta w_{ij}&= -\eta \beta (\langle v_i h_j \rangle_D - \langle v_i h_j \rangle_\lambda),
\end{align}
where the averages above are evaluated over a sample of the data (usually called a minibatch). Although the expressions for the learning rules (\ref{CDapp}) are simple, the computation of (\ref{averagemodel}) is unfeasible in most architectures, since it would involve the knowledge of the partition function, $Z({\beta, \lambda})$, which is a sum of $2^{m\cdot n}$ Boltzmann terms. To avoid this problem, a Markov chain Monte Carlo (MCMC) method can be used to sample the equilibrium distribution by performing the Gibbs sampling dynamics (\ref{dynamics}) for a very large number of iterations. However, the large number of iterations makes the algorithm very slow for practical use in big architectures.

In this sense, contrastive divergence (CD) \cite{Hinton2002} is an idea that simplified the MCMC approach as it approximates the average in (\ref{derivative}) by $\text{n}$ Gibbs steps drawn from the training data using the dynamics (\ref{dynamics}), where the most simple case is $\text{n}=1$. Therefore, the learning rule (\ref{SGA}) for $CD_\text{n}$ is given by
\begin{equation}\label{SGA2}
    \theta_{\tau+1} = \theta_{\tau} + \eta\cdot\beta(\langle \partial_{\theta} E\rangle_{D}-\langle \partial_{\theta}E\rangle_\text{n}),
\end{equation}
In the data average, $\langle \rangle_{D}$, a sample of visible vectors, $v \in D$, is used to generate the hidden vectors, $h$, using (\ref{conditional1}), and $\langle f(v,h) \rangle_D$ is evaluated in the resulting ensemble of states $\{s=(v,h)\}$. Alternatively, the model average $\langle f(v,h) \rangle_{\text{n}}$ represents the empirical average of the function $f(v,h)$ after the application of a $\text{n}$ steps Gibbs sampling from the dynamics (\ref{dynamics}). In this case, a visible vector $v'$ is generated from the hidden vector $h$ using (\ref{conditional2}) and a new hidden vector $h'$ is generated from $v'$ analogously using (\ref{conditional1}). The process is repeated iteratively for $\text{n}$ steps. Finally, $\langle f(v',h') \rangle_\text{n}$ is evaluated as an average in the final ensemble of states of the type $\{s'=(v',h')\}$.

\subsection{Stochastic Thermodynamics of CD}\label{ssCD2}

In the subsection above, it was argued that maximum likelihood (ML) learning deals with the maximization of a known functional (\ref{loglike}), but the gradient ascent steps (\ref{SGA}) require unfeasible computation. Contrastive divergence rule, $CD_\text{n}$, solves this issue by approximating the model average (\ref{SGA2}), although this approximation does not ensure it maximizes the log likelihood (\ref{loglike}). Actually, $CD_\text{n}$ learning \cite{Hinton2002} is equivalent to the minimization of the the following expression:
\begin{equation}\label{CDn}
    CD_\text{n}=KL(p_D\|p_\lambda)-KL(p_n\|p_\lambda),
\end{equation}
where $p_D$ is the data distribution, $p_\text{n}$ is the resulting distribution after $\text{n}$ Gibbs steps (\ref{dynamics}) and $p_\lambda$ is the model distribution (often written as $p_\infty$). The functional $KL(p\|q)$ is the Kullback-Leibler divergence, defined as
\begin{equation}\label{KL}
    KL(p\|q)=\sum_{s}p(s)log\frac{p(s)}{q(s)},
\end{equation}
for probability distributions $p$ and $q$, summed over all states $s$. 
Now we show that functional (\ref{CDn}) has sound physical interpretation based on the stochastic thermodynamics presented in Sec. \ref{maths}. First, notice that the probability $p_\lambda$ is known (\ref{probability}), but there are not closed formulas for $p_D$ and $p_{\text{n}}$. Upon replacing (\ref{probability}) in (\ref{CDn}) and using definition (\ref{KL}) one obtains
\begin{eqnarray}
CD_\text{n}=-\beta\sum_s p_\text{n}(s)E(s,\lambda)+\beta\sum_s p_D(s) E(s,\lambda) \nonumber \\
-\sum_s p_\text{n}(s)\log p_\text{n}(s) + \sum_s p_D(s)\log p_D(s),
\end{eqnarray}
where the partition function $Z(\beta,\lambda)$ has been conveniently canceled out. From the definition of averages (\ref{averagedata}) and (\ref{averagemodel}) and using the definition of the Shannon entropy (\ref{Entropy}), the expression above is rewritten as 
\begin{equation}\label{CDn2}
    CD_\text{n}=-\beta(\langle E(s,\lambda)\rangle_{\text{n}}-\langle E(s,\lambda)\rangle_{D})+ S_\text{\text{n}} - S_0.
\end{equation}
Notice that from the first term above can be written in terms of the stochastic heat defined in (\ref{Heat}) for a constant $\lambda$ and $\text{n}$ steps:
\begin{eqnarray}\label{Heatcte}
Q_\text{n}=\sum_{k=1}^n E(s_{k+1},\lambda)-E(s_k,\lambda) \nonumber\\
=E(s_n,\lambda)-E(s_0,\lambda),
\end{eqnarray}
which in turn allows one to write (\ref{CDn2}) as
\begin{equation}\label{CDnfinal}
    CD_\text{n}= (S_\text{n} - S_0) -\beta \langle Q_\text{n}\rangle.
\end{equation}
The derivation above shows that the $CD_\text{n}$ functional (\ref{CDn}) is composed of two terms. The first term is the variation of the Shannon entropy from the data distribution, $p_D$, to the nonequilibrium distribution, $p_\text{n}$. The second term is minus the average heat observed in the process of taking a data vector and subjecting it to $n$ steps in the Gibbs sampling dynamics (\ref{dynamics}). If the model distribution, $p_\lambda$, is close to the data distribution, $p_D$, their Shannon entropy difference is expected to be negligible, as well as the average heat observed in the process. 
Actually, the nonequilibrium expression (\ref{CDnfinal}) is a measure of how irreversible is this process. The expression turns to a familiar form in the particular case of data being drawn from a MB distribution with configuration $\lambda_D$. By letting $n\rightarrow\infty$, the entropy difference is given by (\ref{EntropyVar}), where $\lambda_0=\lambda_D$ and $\lambda_K=\lambda$. In this case, (\ref{CDnfinal}) becomes the irreversible work
\begin{equation}\label{CDnlast}
    CD_\infty=\beta\langle W \rangle -\beta \Delta F, 
\end{equation}
which is always positive. In other words, contrastive divergence (unsupervised learning) is approximately minimizing the difference between the entropy variation and the average heat (\ref{CDnfinal}) in the process of taking a data vector and placing it in the RBM dynamics. The expression for the optimized functional, $CD_\text{n}$, is well defined in the nonequilibrium stochastic thermodynamics framework. When the number of steps is very large ($n\rightarrow\infty$) and data comes from a MB distribution, the stochastic thermodynamics expression turns to the familiar irreversible work (\ref{CDnlast}), where the partition functions (and the free energy) may be defined.

\section{Application in estimation of the Partition Function}
\label{AIS}
In this section, we show that the Jarzynski Equality (JE) can be explored to estimate the partition function of RBMs with large architectures.

Computing the partition function, $Z=Z(\beta,\lambda)$, of a RBM model (\ref{probability}) is necessary to find the probability of each state according to model. This is important for calculating the Log likelihood (\ref{loglike}) over a data set in order to estimate the performance of a trained model in the unsupervised learning task. However, most practical applications of RBMs uses architectures with large visible and hidden layers \cite{HintonScience2006}, which makes the computation of $Z(\beta,\lambda)$ unfeasible (as a sum of $2^{m\cdot n}$ Boltzmann factors). Different methods have been proposed to estimate the partition function in RBMs in the recent years \cite{AISforRBM,TrackingZBengio}. 

The Annealed Importance Sampling (AIS) \cite{AISNEAL} is a general method for estimating the expectation of some random variable $x$ (or a function of it) drawn from some (intractable) distribution $p(x)$. The idea is based on making a convenient sequence of intermediate distributions that converges to $p(x)$. AIS found application in the estimation of the partition function of RBMs \cite{AISforRBM} with an excellent performance. In the original AIS formalism for RBMs, one defines $p_{\lambda}(v)=p^{*}_{\lambda}(v)/Z(\beta,\lambda)$, where $p_{\lambda}(v)=\sum_{h}p_{\lambda}(s=(v,h))$, obtained from (\ref{probability}), so the estimate of $Z(\beta,\lambda)$, for a configuration $\lambda=(a_i,b_j,w_{ij})$,  can be written in terms of a known partition function $Z(\beta,\lambda_0)$ as:
\begin{equation}\label{AISoriginal}
   \frac{Z(\beta,\lambda)}{Z(\beta,\lambda_0)}
   %=\frac{\sum_v p^{*}_{\lambda}(v)}{Z(\beta,\lambda_0)}
   =\sum_v \frac{p_\lambda^*(v)}{p_{\lambda_0}^*(v)}p_{\lambda_0}(v)
   =\Big\langle \frac{p^{*}_{\lambda}(v)}{p^{*}_{\lambda_0}(v)}\Big\rangle_{p_{\lambda_0}},
\end{equation}
where we used $Z(\beta,\lambda)=\sum_{v}p_{\lambda}^*(v)$ and  $Z(\beta,\lambda_0)^{-1}=p_{\lambda_0}(v)/p_{\lambda_0}^*(v)$ for any $v$. The known configuration $\lambda_0$ could be, for instance, the case $\lambda_0=(a_i,b_j,0)$, for which the partition function can be computed analytically due to its separability (lack of interaction terms between the layers). In the last identity of (\ref{AISoriginal}), we could slightly modify the original AIS formalism in order to sum over all possible states of the RBM, $s=(v,h)$, for a clearer interpretation within thermodynamics. It results in an equivalent expression
\begin{eqnarray}\label{AISnew}
   \frac{Z(\beta,\lambda)}{Z(\beta,\lambda_0)}
   =\sum_s p_{\lambda_0}(s)e^{-\beta (E(s,\lambda)-E(s,\lambda_0))},
\end{eqnarray}
where the definition $Z(\beta,\lambda)=\sum_s e^{-\beta E(s,\lambda)}$ was used, as well as the identity $Z(\beta,\lambda_0)^{-1}=p_{\lambda_0}(s)e^{\beta E(s,\lambda_0)}$ that comes from (\ref{probability}). One can easily notice that the exponents in (\ref{AISnew}) are the stochastic work (\ref{Work}) defined in the thermodynamics formalism for a system prepared at equilibrium ($\beta,\lambda_0$) after a single ($k=1$) Gibbs step (\ref{dynamics}) with parameters ($\beta,\lambda$). So it is immediate that one could estimate $Z(\beta,\lambda)$ by computing the average of a stochastic quantity $e^{-\beta W}$, a function of the stochastic work, $W$, over a ensemble of states starting from a known MB distribution with configuration ($\beta,\lambda_0$). However, the quality of such estimate depends on the size of the ensemble, and since $\lambda_0$ and $\lambda$ may differ greatly, the variance of sampling (\ref{AISnew}) (the single step protocol) may be large.

Fortunately, a refinement of this single step estimate can be done by considering a slow protocol going from the configuration $(\beta, \lambda_0)$ to $(\beta, \lambda=\lambda_K)$ as sequence of intermediate steps $\lambda_k$. Notice that, using (\ref{AISnew}), the ratio between partition functions $Z(\beta, \lambda_K)$ and $Z(\beta, \lambda_0)$ can be written as the product
\begin{eqnarray}\label{productZ}
   \frac{Z(\beta,\lambda_K)}{Z(\beta,\lambda_0)}=\prod_{k=0}^{K-1} \frac{Z(\beta,\lambda_{k+1})}{Z(\beta,\lambda_{k})}=\langle e^{-\beta W} \rangle,
\end{eqnarray}
since the intermediate factors cancel out in the first identity above. The last identity follows from using (\ref{AISnew}) in each factor of the product. In this expression, the average is taken over all possible trajectories $\Sigma=(s_0,...,s_K)$, with the weights being adjusted in a controllable protocol, $\Lambda=(\lambda_0,...,\lambda_K)$. Notice that the ratio of the partition functions in (\ref{productZ}) may be written in terms of the free energy, $\beta\Delta F = -\log(Z(\beta,\lambda)/Z(\beta,\lambda_0))$, which makes the expression equivalent to the Jarzynski equality (\ref{JE}), obtained independently \cite{Jar1997} from AIS. This evidence supports the claim that AIS and JE provides essentially the same method for computing the partition function, as claimed originally in \cite{AISNEAL}. 

A close inspection in (\ref{AISnew}) shows that it could only be used in the step $k$ of the protocol $\Lambda$ in (\ref{productZ}) if the system is approximately in equilibrium in the configuration $(\beta,\lambda_{k-1})$. Therefore, the protocol $\Lambda$ should be slow enough ($||\lambda_{k}-\lambda_{k-1}||\ll1$) to account for this condition. Physically, this protocol represents a quasi-static (reversible) isothermal ``expansion'' (or ``compression''), where the configuration $\lambda$ could be understood as an external set of controlled parameters, akin to the volume of the system in equilibrium thermodynamics. By performing a slow protocol, the system is managed to stay locally in thermal equilibrium, ie, its nonequilibrium distributions are approximately MB during the whole process.  

Actually, the physical condition of a quasi static process $\Lambda$ is met in real AIS estimations of the partition function \cite{AISforRBM}. Typically, simulations regarding AIS uses a specific time protocol $\lambda_t=(a_i,b_i,w_{ij}(t))$, where the weights $a_i$ and $b_i$ are constant and the interaction terms $w_{ij}(t)$ goes from $0$ to $w_{ij}$ in a linear or exponential behavior. Due to the functional form of the MB distribution (\ref{probability}), this sort of transformation in the the configuration parameters resembles a fine tuning in the inverse temperature of the system. Therefore the process is often seen as a simulated annealing approach. However, we point out that the work protocol interpretation of the AIS approach presented in this section (\ref{productZ}), in which the temperature is held constant, corroborates with the stochastic thermodynamics framework of RBMs.

\section{Summary and conclusions}
\label{conclude}

In this paper, we have analyzed Restricted Boltzmann Machines (RBMs) in a stochastic thermodynamics approach. We start by presenting the RBM as discrete Markov chains satisfying detailed balance condition. This property allowed us to  adapt the framework \cite{Crooks1998} to define the stochastic heat and work, leading to the first law of thermodynamics.

We highlighted nonequilibrium fluctuation theorems arising from this approach. Notably, the Crooks Fluctuation Theorem (CFT) followed immediately, since it was originally derived in the context of Markov chains. The Jarzynski equality (JE) and the second law of thermodynamics followed from CFT as expected. Then, the heat exchange fluctuation theorem (XFT) was derived for any configuration of RBMs, which differs from its original presentation \cite{Jar2004} based on a hamiltonian with a small thermal coupling. Our presentation uses the general facts that the equilibrium distribution is MB (\ref{probability}) and the detailed balance condition. Numerical simulations in a fully trained RBM shows excellent agreement with XFT predictions in the nonequilibrium case of a single Gibbs step in the dynamics, $K=1$, for both heating and cooling situations.

We also interpreted the known contrastive divergence (CD) unsupervised learning algorithm \cite{Hinton2002} in the context of stochastic thermodynamics. We showed that the $CD_n$ functional of two distributions ($p,q$), defined in terms of the Kullback-Leibler divergence, can be written as thermodynamic observables. It turns out that $CD_\text{n}$ is a measure of how irreversible is the process of propagating data vectors with the RBM dynamics. Namely, the $CD_\text{n}$ functional to be optimized in the learning process is the difference between the entropy variation and the average stochastic heat of that process. In the particular case of MB distributions, for a infinite number of steps, the expression is reduced to the known irreversible work, which plays important role in stochastic thermodynamics \cite{Sekimoto2010}.

Finally, we presented how the ratio of partition functions may be estimated by averaging a thermodynamic observable. The derivation, that is closely related to the Jarzynski equality (JE), is mathematically equivalent to the widely used Annealed Importance Sampling (AIS) algorithm, as claimed in the original derivation of AIS \cite{AISNEAL}. The difference of interpretations being that, in the stochastic thermodynamics framework, a work protocol is produced in the estimation of the partition function at constant temperature. A process that resembles a physical isothermal transformation, opposed to the original annealing interpretation in which the temperature is slowly changed during the process.

We point that artificial neural networks (ANNs) have produced astonishing results over the years for image, text and speech recognition, specially when stacked in form of multiple layers, known as deep learning. Most of the applications, including the ones observed in physics \cite{Atlas1,Atlas2,PhaseTrans2016} are trained supervised, a situation that requires a lot of labeled data. Actually, the vast majority of available data in the world is not labeled. Biologic systems also learn representations of the world from sensory data in a unsupervised manner. These observations make generative models to be speculated as the next frontier in artificial intelligence, for which the Restricted Boltzmann Machines are a type of building block. The results presented in this paper supports that unsupervised learning obey general rules observed in thermodynamics, mostly due to the fundamental properties of its dynamics, such as detailed balance. The results are also general enough and could possibly be extended to deep ANNs such as the Deep Belief Network (DBN). 
\section{Acknowledgments}
\label{acknowledgments}
This work was supported by Funda\c{c}\~ao de Amparo \`a Ci\^encia e Tecnologia de Pernambuco (FACEPE) under grant APQ $0073$-$1.05/15$.
\appendix
\label{Appendix1}
\section{Experiments}
In this section, the known benchmark MNIST data set is briefly described. We also provide details on the unsupervised learning experiment on MNIST that produced the heat exchange fluctuation theorem (XFT) of subsection \ref{XFTsec}.
\subsection{MNIST data set}
The MNIST is a data set of images of handwritten digits of size $28\times28$ pixels \cite{MNIST}. It is widely used as a benchmark for machine learning algorithms. The set contains $60,000$ images used for training and $10,000$ images used for testing. The images were binarized, so pixel values are either 0 or 1. Since its creation, several algorithms have reached very low error rates for the supervised (or labeled) problem, aimed to classify an image in one of the $10$ categories (digits). However, the goal of section (\ref{XFTsec}), as in previous unsupervised learning applications of RBMs \cite{Hinton2002,Hinton2006}, is to learn the handwritten data distribution by trying to optimize the log likelihood (\ref{loglike}). It means the algorithm should try to generate the original distribution of MNIST images as close as possible to the original set, without using any information of the image labels.

\subsection{Training the RBM}
The training procedure for the contrastive divergence algorithm is straightforward. We use the increments for the parameters $\lambda=(a_i,b_j,w_{ij})$ from (\ref{CDapp}). It is worth to point out that the average in $D$ is to be understood as taken from data (positive phase) and the average $\text{n}=1$ is taken from the reconstructed image (negative phase) \cite{Hinton2002}, after a single Gibbs sampling starting from the original image. 
The RBM has the same architecture with $m=784$ (representing $28\times28$ pixels of MNIST) and $n=500$ neurons in the visible and hidden layers, respectively. We initialize the biases ($a_i,b_j$) at zero and $w_{ij}$ from a uniform distribution (from $-0.1$ to $0.1$).
The training set is split in $600$ minibatches of $100$ images. For each iteration, all the images of a minibatch are used to generate the positive and negative phases, used to compute the increments of the weights (\ref{CDapp}) and the weights are updated. Passing through all minibatches is called an epoch. In our experiment, the learning rate was set $\eta=0.004$ for $300$ epochs. A linear weight decay of $\alpha=10^{-4}$ was used and momentum was set to $0$. The inverse temperature is a constant $\beta_1=1$. The parameters $\lambda$ of the network were trained using a slight modification of CD, called the Persistent Contrastive Divergence (PCD) algorithm \cite{PCD} in which the positive phase ensemble in (\ref{CDapp}) is not restarted from data, but taken from a persistent value reused from last epoch. This simple modification yields better results in the data generation task without increasing computational cost.

\end{document}